\documentclass[%
 reprint,
 amsmath,amssymb,
 aps,
prstab,
]{revtex4-1}

\usepackage{lipsum} 
\usepackage{color}
\usepackage{siunitx}
\usepackage{amsmath,amsfonts,amsthm}
\usepackage{graphicx, subfigure}
\usepackage{placeins}
\usepackage{float}
\usepackage{subfigure}
\usepackage{verbatim}
\usepackage{dirtytalk}
\usepackage{comment}
\begin{document}
\newcommand{\etal}{\textit{et al}. }


\title{Nonlinear wakefields and electron injection in cluster plasma}


\author{M. W. Mayr$^1$}
\email{marko.mayr@physics.ox.ac.uk}
\author{B. Spiers$^1$}
\author{R. Aboushelbaya$^1$}
\author{R. Paddock$^1$}
\author{J. D. Sadler$^2$}
\author{C. Sillett$^1$}
\author{R. H. W. Wang$^1$}
\author{K. Krushelnick$^3$}
\author{P. A. Norreys$^{1,4,5}$}

\affiliation{$^1$Clarendon Laboratory, University of Oxford, 
Oxford OX1 3PU, UK}
\affiliation{$^2$Los Alamos National Laboratory, P.O.Box 1663, Los Alamos, New Mexico 87545, USA}
\affiliation{$^3$Center for Ultrafast Optical Science, University of Michigan,
Ann Arbor, MI, USA}
\affiliation{$^4$Central Laser Facility, STFC, Rutherford Appleton Laboratory,
Didcot OX11 0QX, UK}
\affiliation{$^5$John Adams Institute, Denys Wilkinson Building, Oxford OX1 3RH, UK}



\date{\today}

\begin{abstract}
Laser and beam driven wakefields promise orders of magnitude increases in electric field gradients for particle accelerators for future applications. Key areas to explore include the emittance properties of the generated beams and overcoming the dephasing limit in the plasma. In this paper, the first in-depth study of the self-injection mechanism into wakefield structures from non-homogeneous cluster plasmas is provided using high-resolution two dimensional particle-in-cell simulations. The clusters which are typical structures caused by ejection of gases from a high-pressure gas jet have a diameter much smaller than the laser wavelength. Conclusive evidence is provided for the underlying mechanism that leads to particle trapping, comparing uniform and cluster plasma cases. The accelerated electron beam properties are found to be tunable by changing the cluster parameters. The mechanism explains enhanced beam charge paired with large transverse momentum and energy which has implications for the betatron x-ray flux. Finally, the impact of clusters on the high-power laser propagation behavior is discussed.


\end{abstract}


\maketitle

 
 \section{INTRODUCTION}
A cluster plasma is a collection of ionized atomic clusters of various sizes which are usually randomly distributed. These clusters can either make up the entirety of the plasma or are embedded within a uniform plasma. Cluster plasmas are produced when atomic structures which were formerly held together by van der Waals forces are ionized within a short amount of time. The clusterization process in high pressure gas jets was described and characterized in \cite{Hagena,Smith,Murakami,Lu2010, Grieser, Tao, Aladi_2017}.

The interactions between short laser pulses and high-density clusters has been studied extensively \cite{Ditmire,breizman,fennel} due to the variety of applications from sources of soft x-ray or EUV emission$\,$\cite{McPherson1994,Skobelev2002,Fukuda08} to high harmonic generation \cite{Donnelly,Tisch_1997,Aladi_2014,Tao_2017}. These articles found that differences in cluster density, cluster diameter, laser amplitude, or electric field gradient have a significant impact on the expansion of the cluster. For high-power laser pulses, depending on the aforementioned cluster parameters, the laser field either causes electrons to be driven back into the cluster during each laser cycle leading to resonant heating and efficient laser pulse absorption \cite{antonsen, taguchi, mulser, zamith} or causes a Coulomb explosion through removing a great majority of electrons from the ion cluster \cite{esirkepov,teuber}. The optical properties within the plasma can be changed by electrons bound to a cluster structure. A laser pulse which propagates through a cluster medium can access the ``cluster mode"$\,$\cite{tajima2}. This allows it to propagate in a plasma containing clusters of highly overcritical density at a group velocity higher than that in a uniform plasma of the same average density.

Wakefields are electron plasma waves driven by a high power laser pulse or a relativistic particle bunch. They are able to generate accelerating (and focussing) electric fields of up to $\SI{100}{\giga\volt\per\meter}$. Through the theoretical study of the ``blowout" or ``bubble"$\,$\cite{Pukhov2002,Pukhov2004,Kostyukov2004,Lu2006} regimes, efficient acceleration schemes using electron self-injection were predicted that lead to the production of quasi-monoenergetic beams. A phenomenological model using observables such as the wake amplitude, laser depletion, and beam loading as physical parameters was developed by Lu \etal$\,$\cite{Lu2007}. The authors showed for normalized laser amplitudes of $a_0=eA/(m_ec^2)\geq 2$ and pulse lengths $\tau$ shorter than $w_0/c \simeq 2\sqrt{a_0}/\omega_p$ that the laser spot-size $w_0$, wakefield ``bubble'' radius $R$, and normalized laser amplitude $a_0$ are linked through $k_pR\simeq k_pw_0 = 2\sqrt{a_0}$ and put it into relation to the critical power for self-focusing $P_c = 17 \omega_0^2/\omega_p^2 [\si{\giga\watt}]$ found by Sun \etal$\,$\cite{Sun1987}. Here, $e$ is the elementary charge, $A$ is the laser vector potential, $m_e$ is the electron rest mass, $c$ is the speed of light in vacuum, $\omega_p$ is the plasma frequency, and $\omega_0$ is the laser frequency. 
Laser-driven wakefields have proven to be able to accelerate high-charge electron beams of high quality in$\,$\cite{Mangles2004,Geddes2004,Faure2004,Leemans2006}. Following these milestone experiments the race for higher particle energies and better beam quality has produced promising results$\,$\cite{Lundh2011,Mo2012,Wang2013, Kim2013,Leemans2014, Gonsalves2019}.

The study of cluster plasmas is relevant for all experimental studies involving high pressure gas jets and thus also for plasma wakefield accelerator experiments. The behavior of wakefields in clustered plasmas deviates from the standard uniform plasma case as was shown for a non-relativistic laser amplitude in \cite{PhysRevAccelBeams.22.113501}. Experimental campaigns have looked at accelerated electrons from clustering gas targets (using methane and argon) discovering an impact on electron beam features due to the presence of clusters. These experiments observed a broader energy spectrum of the accelerated electron beam \cite{Fukuda,Zhang2012,Chen2013,Mirzaie2016,Woodthesis,Dann2019}, increased amounts of accelerated charge \cite{Chen2013,Mirzaie2016}, higher peak energies \cite{Chen2013,Mirzaie2016}, and better reproducibility \cite{Woodthesis} using cluster targets. Researchers argued (\cite{Fukuda,Chen2013}) that a significant part of the acceleration of cluster electrons is due to direct laser acceleration (DLA). It has to be noted that their interpretations rely on simulations that either study much larger clusters of $\SI{0.24}{\micro\meter}$  in diameter \cite{Fukuda} or use insufficient resolution of 1 cell per cluster diameter using 9 particles per cell \cite{Chen2013}. 

In this article, we investigate how the presence of clusters leads to a new injection mechanism featuring deviations in properties such as charge, particle beam energy, and emittance. Furthermore, we investigate how different cluster parameters might be able to increase the dephasing length through the clusters' impact on the group velocity of the short laser pulse propagating through the plasma. 

This article is structured as follows. Section \ref{sec:wakefield_behavior} elaborates on the wakefield properties in a cluster plasma while section \ref{sec:acceleration} looks at the self-trapping mechanism and shows how cluster-born electrons are accelerated in a cluster plasma. Section \ref{sec:laser} shows how the laser propagation is impacted with varying cluster parameters. This is followed by a discussion of the results presented.

\section{BLOWOUT REGIME FOR CLUSTER PLASMA}\label{sec:self_trapping}
\subsection{Simulation Parameters}
In order to investigate the wakefield behaviour at high laser amplitudes in a cluster plasma, we conducted two-dimensional moving-window particle-in-cell (PIC) simulations using the fully relativistic OSIRIS code$\,$\cite{OSIRIS}. The cell-size within the cluster simulations had to be very small to fully resolve nanometer-scale clusters - a length scale which is much smaller than the laser wavelength in typical high-power laser experiments. The simulation grid for the simulation with highest resolution (smallest cluster diameter) was made of 15000$\times$8400 cells which corresponds to 188.5 cells per wavelength in the longitudinal (x) direction and 164.9 cells per wavelength in the transverse (y) direction. The high spatial resolution leads to very accurate laser numerical dispersion of the electromagnetic wave with the Maxwell solver (in this work, a Yee solver was used) and reduces artificial numerical heating in the clusters. The uniform plasma simulations had lower resolution requirements and used 37.7 cells per wavelength in the longitudinal (x) direction and 12.6 cells per wavelength in the transverse (y) direction (a resolution high enough to prevent numerical dispersion).  
The electron density averaged over the entire simulation box used for most simulations was $\langle n_e\rangle =0.004\,n_\text{cr}$, corresponding to $6.97\times 10^{18}\si{\per\centi\meter\cubed}$, where $n_\text{cr} = \omega_0^2m_e\epsilon_0/e^2$ is the critical density of plasma for the drive pulse frequency $\omega_0$ corresponding to a laser wavelength of $\SI{800}{\nano\meter}$. The clusters are injected pre-ionized at an electron temperature of $\SI{10}{\electronvolt}$ and using cold ions as it was found in preliminary simulations that ionization had a negligible impact on the results discussed in what follows.  \newline
The normalized laser amplitude, $a_0$, was set to be 5.0 at the focal position where $a_0\equiv8.55\times10^{-10} (I\lambda^2 [\text{W\,cm}^{-2}\mu\text{m}^{2}])^{1/2}$ ($I$ is the intensity in $\si{\watt\per\centi\meter\squared}$ and $\lambda$ is the laser wavelength in $\si{\micro\meter}$) while the spot size and duration was varied for different simulations. This amplitude corresponds to a peak intensity of $I = \SI{5.34e19}{\watt\per\centi\meter\squared}$. The laser had linear p-polarization with a Gaussian (longitudinal and transverse) shape. 
There were $32\times32$ electron macro-particles per cell for the argon simulations, $16\times16$ electron macro-particles per cell for the hydrogen simulations and $8\times8$ ion macro-particles per cell in all cluster simulations. The particle boundaries were absorbing and the longitudinal (transverse) field boundaries were of type ``open'' (``Lindman''$\,$\cite{LINDMAN197566}). The simulation time-step was chosen to satisfy the Courant-Friedrichs-Lewy condition$\,$\cite{Courant1928}. The uniform plasma simulations used $8\times8$ electron macro-particles per cell and $2\times2$ ion macro-particles per cell.
All other parameters are shown in table$\,$\ref{tab:sims}.

\begin{table}[h!]
\centering
 \begin{tabular}{||c| c| c| c|c|c|c|c|c||} 
 \hline
 \# & Type & $P_\mathrm{L}$ & $d$ & $w$ & $n_\mathrm{cl}/n_\mathrm{cr}$ & $d_\mathrm{cl}$ & Gas & $\langle n_e\rangle/n_\mathrm{cr}$ \\ [0.5ex] 
 \hline\hline
   & - & [$\si{\tera\watt}$] & [$\si{\femto\second}$] & [$\si{\micro\meter}$] & - & [$\si{\nano\meter}$] & - & -\\ [0.5ex]  
 \hline\hline
 A & unif. & 196  & 30 & 12.7 & - & - & Ar$^{16+}$ & 0.004\\ [0.5ex]  
 \hline
 B & cl.(s) & 196 & 30 & 12.7 & 1.5 & 12 & Ar$^{16+}$ & 0.004\\[0.5ex] 
 \hline
 C & cl.(r) & 196 & 30 & 12.7 & 1.5 & 12 & Ar$^{16+}$& 0.004\\[0.5ex]  
 \hline\hline 
 D & unif. & 332 & 30 & 16.6 & - & - &H$^{+}$& 0.004\\[0.5ex] 
 \hline
  E & cl.(r) & 332 & 30 & 16.6 & 1.25 & 18 & H$^{+}$& 0.004\\ [0.5ex] 
 \hline
  F & cl.(r) & 332 & 30 & 16.6 & 5.0 & 18 & H$^{+}$& 0.004\\ [0.5ex] 
 \hline
  G & cl.(r) & 332 & 30 & 16.6 & 0.75 & 18 & H$^{+}$& 0.004\\ [0.5ex]  
 \hline
  H & cl.(r) & 332 & 30 & 16.6 & 2.5 & 24 & H$^{+}$& 0.004\\ [0.5ex] 
 \hline\hline
  I & unif. & 332 & 30 & 16.6 & - & - &H$^{+}$& 0.002\\ [0.5ex] 
 \hline
  J & cl.(r) & 332 & 30 & 16.6 & 1.25 & 24 & H$^{+}$& 0.002\\ [0.5ex]  
   \hline\hline
  K & unif. & 126 & 15 & 10.2 & - & - &H$^{+}$& 0.004\\ [0.5ex] 
 \hline
  L & cl.(r) & 126 & 15 & 10.2 & 1.25 & 24 & H$^{+}$& 0.004\\ [0.5ex]  
 \hline
 \end{tabular}
\caption{List of simulation parameters used. The peak normalized laser amplitude for all simulations was $a_0 = 5.0$. $P_L$ is the laser power,  $d = d_\mathrm{FWHM}$ is the laser (intensity) FWHM duration, $w = w_\mathrm{FWHM}$ is the laser (intensity) FWHM-spot diameter, $n_\mathrm{cl}/n_\mathrm{cr}$ is the cluster peak density normalized to the critical density, and $d_\mathrm{cl}$ is the cluster diameter. "unif." indicates uniform density distribution, "cl.(s)" indicates the use of periodically arranged clusters defined by an analytic function, and "cl.(r)" indicates the use of clusters of randomized positions.}
\label{tab:sims}
\end{table}

As a reference, cluster sizes for argon and hydrogen gas jets were studied for different geometries in \cite{Grieser, Aladi_2017} finding cluster radii between $\SI{8}{\nano\meter}$ and $\SI{20}{\nano\meter}$ for backing pressures between $\SI{12}{\bar}$ and $\SI{16}{\bar}$.


\subsection{Wakefield Behavior}\label{sec:wakefield_behavior}
Here, we show how replacing a uniform plasma with ionized clusters changes the behavior of the nonlinear wakefield in the ``bubble regime"$\,$\cite{Pukhov2002,Pukhov_2004}. To this end, we discuss the results of the simulations listed in Tab$\,$\ref{tab:sims}.

\begin{figure}
    \includegraphics[width=1\linewidth]{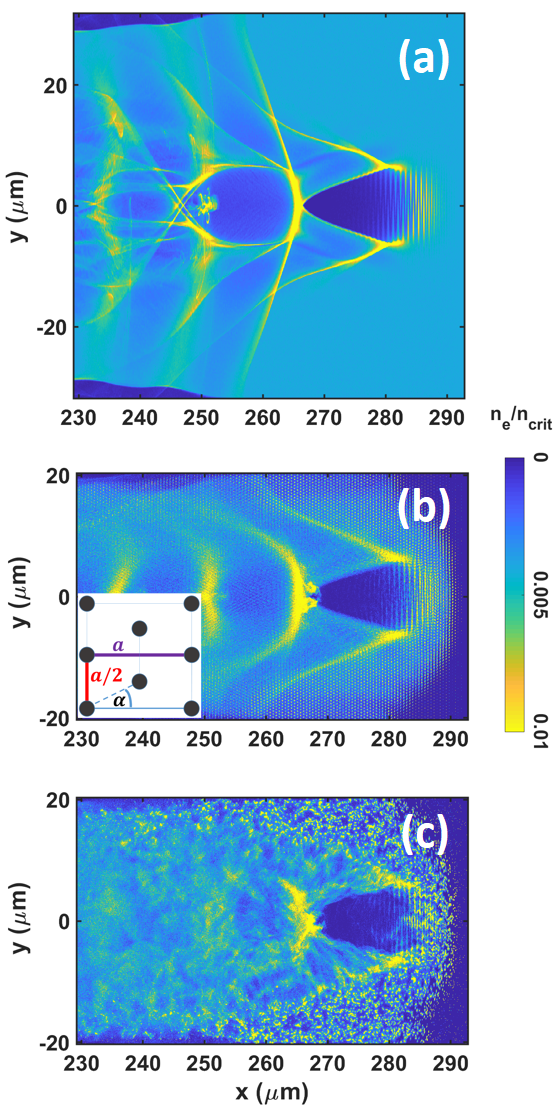}
    \caption{Results of PIC simulations A, B, and C (parameters according to Tab.$\,$\ref{tab:sims}). Electron density normalized to critical density for (a) a uniform plasma, (b) a plasma of clusters arranged in a staggered grid, and (c) a plasma of clusters with random positions after $\SI{0.76}{\pico\second}$. }
    \label{fig:wakefield_comp}
\end{figure}

Fig.$\,$\ref{fig:wakefield_comp} compares the electron charge density of a wakefield driven by a $\SI{30}{\femto\second}$ laser pulse of $a_0=5$ in a uniform plasma (simulation A), a plasma of clusters arranged on a staggered grid (B), and a plasma of randomly distributed clusters (C). 
In the uniform case, the wakefield enters the ``bubble'' regime as expected. A similar wakefield structure can be observed in the case in which the laser pulse is propagating through overcritical clusters arranged in a pattern as shown in the schematic inlay in panel (b). The clusters are staggered at an inclination angle of $\alpha = 26.57^{\circ}$ and the closest distance between cluster centers is $a/2=\SI{246.4}{\nano\meter}$. The bottom panel shows how randomizing cluster locations while keeping all other cluster parameters unchanged deteriorates the bubble symmetry and the well-known wakefield structure. This shows the necessity of enabling cluster randomization in simulations in order to capture important features of the physical picture of the acceleration process which - to our knowledge - has not been done before.

\begin{figure*}
    \includegraphics[width=0.9\linewidth]{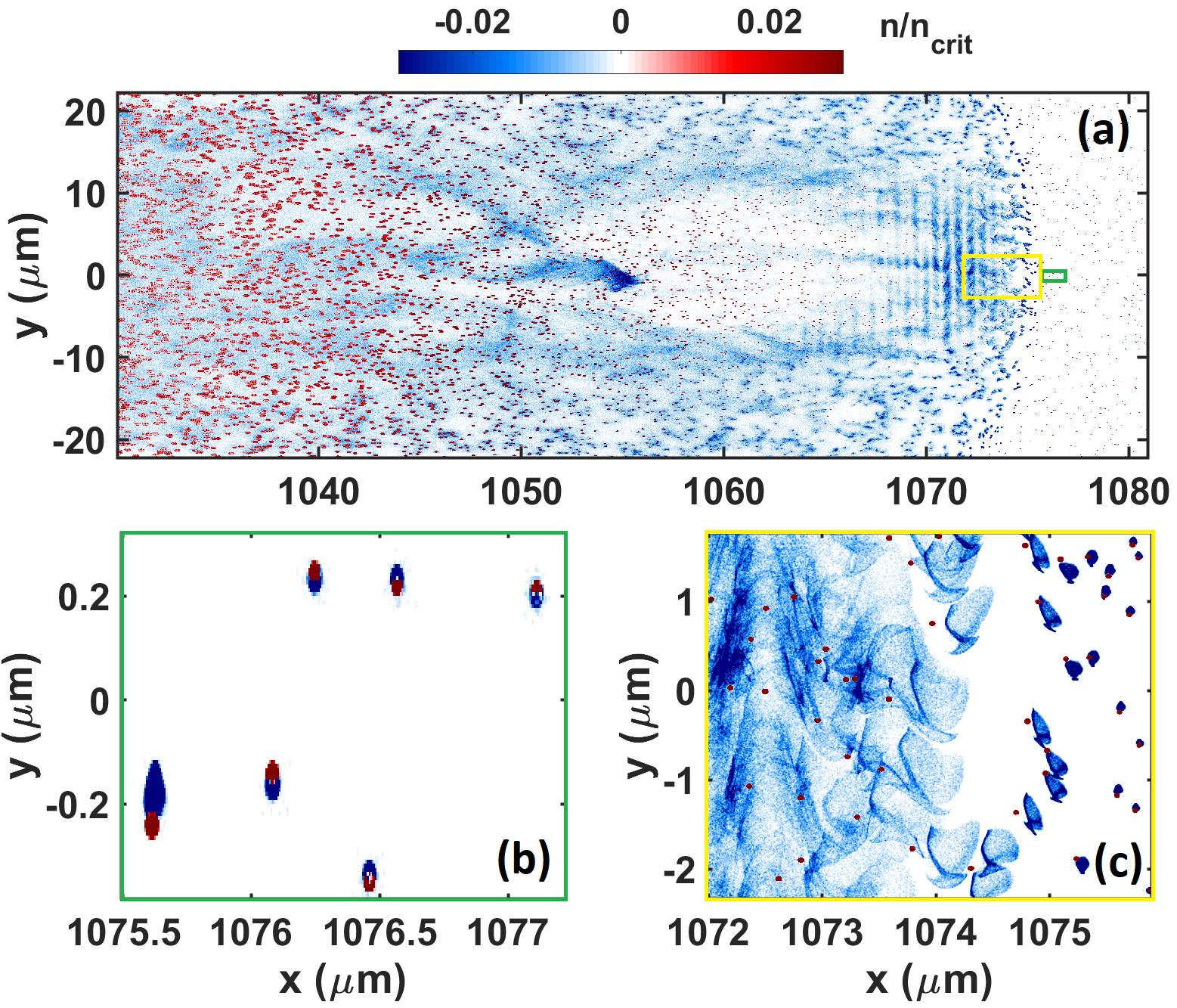}
    \caption{Normalized charge density of simulation (G) after $\SI{3.57}{\pico\second}$. Panel (a) shows the wakefield structure formed by plasma electrons. The bunch of trapped and accelerated electrons can be clearly seen. The ion clusters (red) expand towards the back of the simulation box (left) through the Coulomb explosion mechanism. The green and yellow boxes indicate the areas shown in panels (b) and (c). Panel (b) shows a snapshot of the hydrodynamic expansion of the clusters in the region where the laser electromagnetic fields are of low magnitude. Panel (c) shows the region in which the clusters encounter the ponderomotive force pushing them forwards (towards the right) and away of the ion cluster and the Lorentz force causing an oscillatory motion about the cluster center depending on the sign of the electric field at the respective position. Additionally, the Coulomb explosion lets the electron bunches expand on the order of a few $\si{\micro\meter}$.}
    \label{fig:stream}
\end{figure*}

Following the argument of \cite{Lu2007}, the blowout radius of the wakefield is $R \simeq 2\sqrt{a_0}\lambda_p/(2\pi)$. Generally, it can be said that when the laser is close to focus, the bubble minor and major axes are of similar size in the cluster cases compared to the uniform plasma case in simulations using the same total averaged density. At later timesteps it is more difficult to measure these values for the randomized cluster case due to strong shape fluctuations and asymmetries. Due to the random cluster distribution and corresponding high density regions (clusters of clusters so to say), high-density ``jets" of electrons were observed to ``stream" towards the left into the evacuated region in the co-moving frame in all cluster simulations with randomized cluster position (see Fig.$\,$\ref{fig:stream}). As will be discussed later, this effect aids a new type of electron trapping within the wakefield.  

\begin{figure}
    \centering
    \includegraphics[width=1\linewidth]{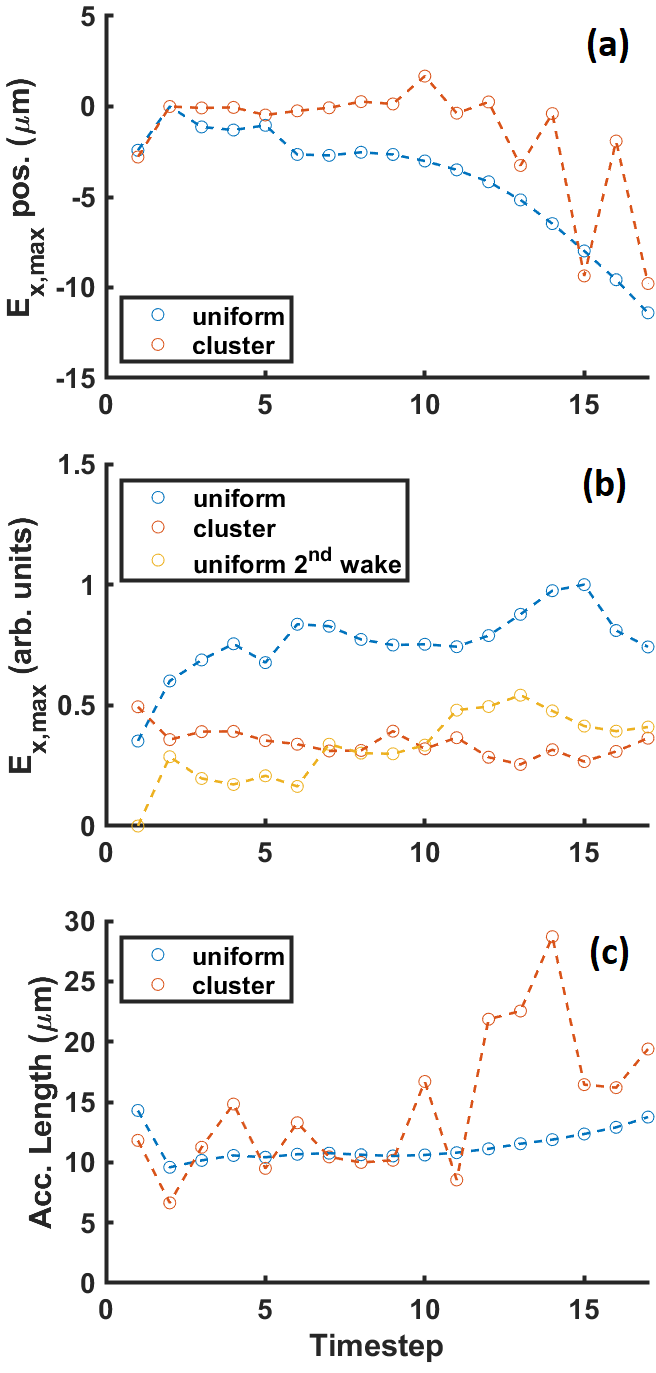}
    \caption{Comparison of the wakefield evolution in a uniform and a cluster plasma (simulations D and E - see table \ref{tab:sims}). Panel (a) shows the position of the maximum accelerating longitudinal field plotted against time. The zero (reference) value on the y-axis was chosen to be at time step 2 where the maximum plasma density has been reached after the density ramp. Panel (b) shows the normalized peak accelerating electric field strength for cluster and uniform cases as well as the value for the second wake in the uniform case. Panel (c) shows the maximum acceleration length, defined as the distance between troughs where the electric field in the first bubble is negative (and thus accelerating). Here, one time step corresponds to $\SI{0.25}{\pico\second}$ in simulation D and to $\SI{0.26}{\pico\second}$ in the other simulations. All values are taken from a line-out on the laser propagation axis. }
    \label{fig:analysis}
\end{figure}

Fig.$\,$\ref{fig:analysis} quantitatively shows how the wakefield behaviour differs when using a cluster plasma instead of a uniform plasma. The top panel compares the position of the maximum longitudinal electric field $E_x$. After propagating through a density ramp, the laser pulse has reached the maximum density region at time step 2 explaining why the maximum electric field position increases up to this point (plasma wavelength decreases approaching higher density). The laser is focused at $\SI{202.45}{\micro\meter}$ which corresponds to a position it has reached shortly before the data output of time step 3. The position slips backward in a stable manner in the uniform case due to the difference between the wakefield phase velocity $v_\Phi=v_g-v_{etch}$ and the simulation box propagation velocity $c$ whilst maintaining a similar bubble radius.  In contrast to this robust accelerating structure which has been studied in \cite{Lu2007}, the cluster simulation shows strong fluctuations of the electric field peak position. 
The magnitude of this electric field can be seen in the middle panel of Fig.$\,$\ref{fig:analysis}. The peak value is consistently lower in the cluster plasma case when comparing the first respective wake to the uniform plasma case. This is due to the imperfect evacuation of electrons from the bubble region in the cluster case and the deviations from the symmetric plasma wave. 
The field strength in the cluster case is more similar to the second uniform wakefield into which we observe particles to be injected over the duration of the simulation. This means that when looking at electron energy gain (see section \ref{sec:acceleration}), we will be comparing acceleration in similar field strengths. 
We define the acceleration length as the longest distance within the wakefield structure in which injected particles could be accelerated - a value obtained from taking the distance between the two troughs in the $E_x$ data that are furthest apart (in this simulation this always corresponds to the first wake). This means we are measuring the on-axis distance over which $E_x$ is negative. The bottom panel in Fig.$\,$\ref{fig:analysis} shows a very stable (slowly increasing) accelerating region in the uniform plasma. This is in strong contrast to the cluster simulation in which the acceleration length fluctuates by more than $\SI{20}{\micro\meter}$. We will show later that this enables particles to be injected at trajectories uncommon for self-injection and that different cluster parameters impact this value. 

\begin{figure}
\subfigure{\label{fig:Exu1}\includegraphics[width=1.0\linewidth]{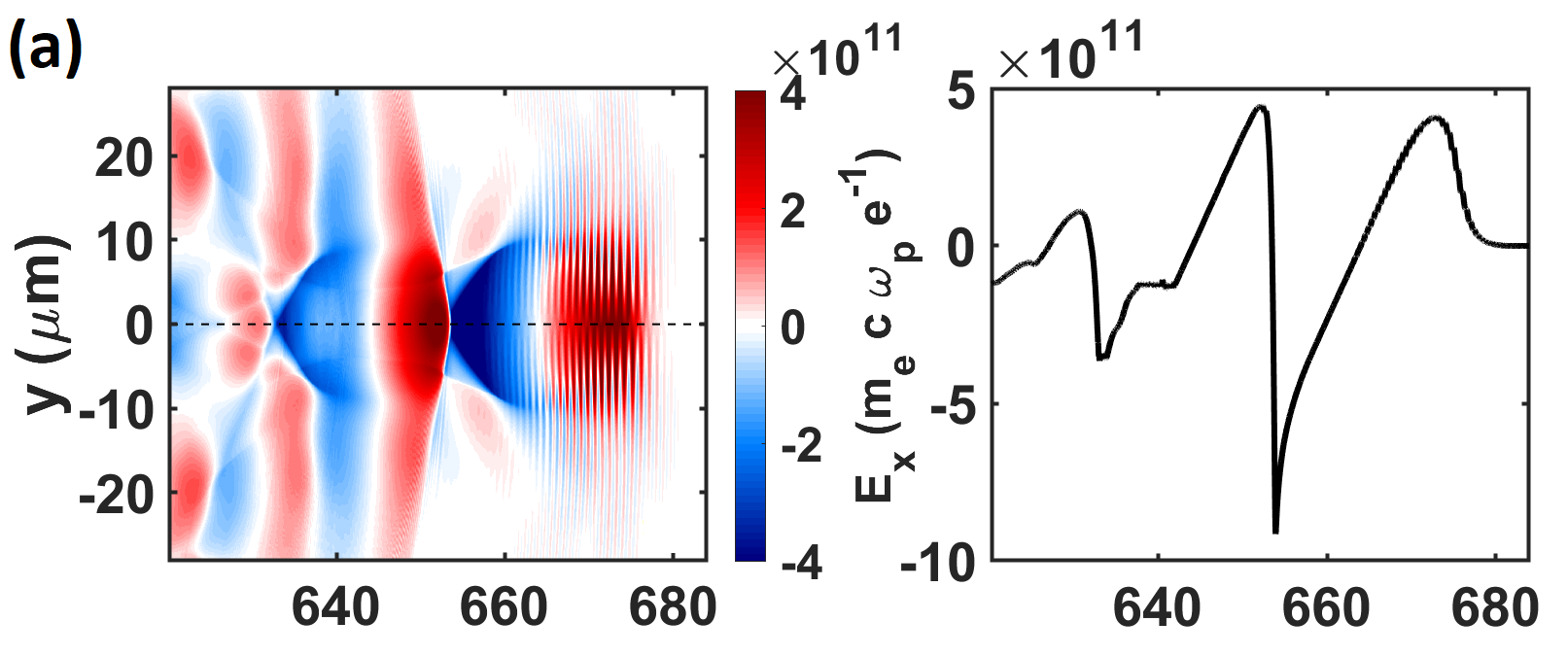}}
\subfigure{\label{fig:Exc1}\includegraphics[width=1.0\linewidth]{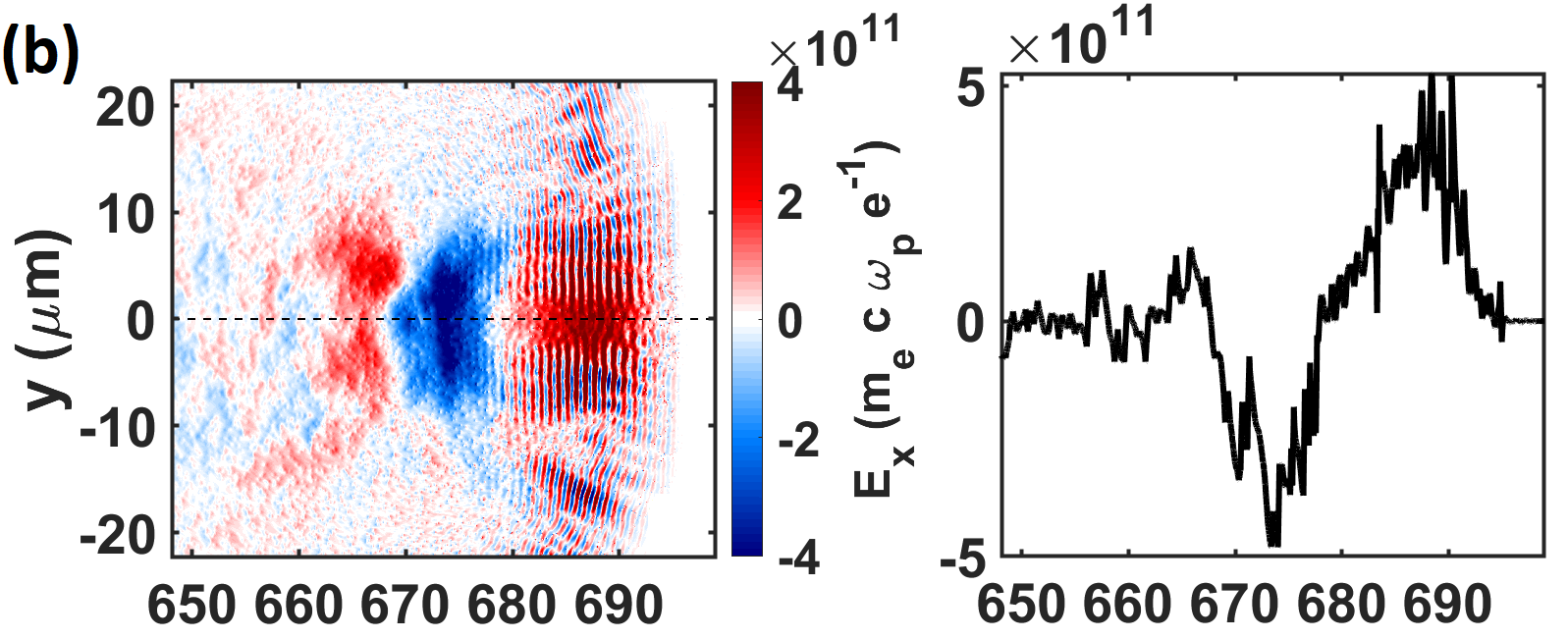}}
\subfigure{\label{fig:Exc5}\includegraphics[width=1.0\linewidth]{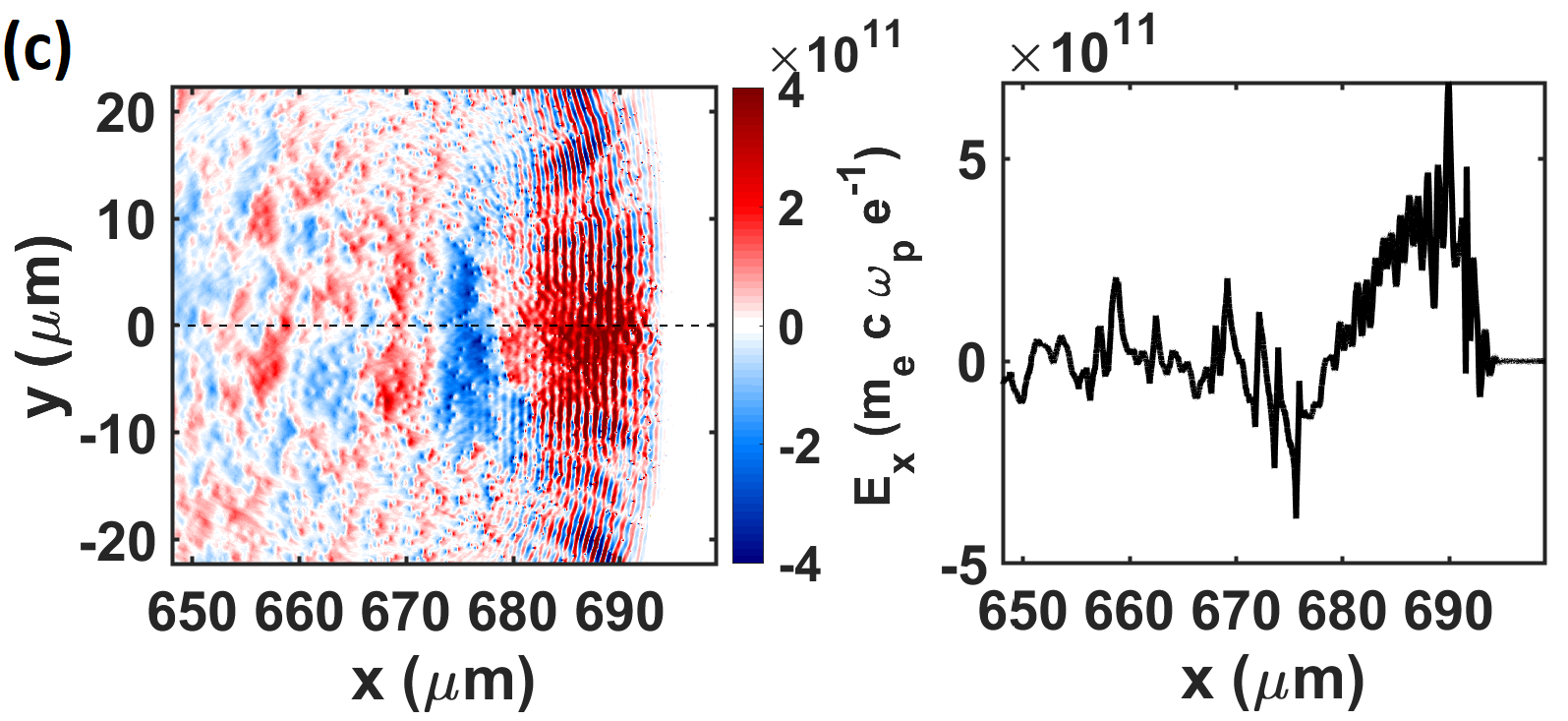}}
\caption{Longitudinal electric field of simulations D (top row), E (middle row), and F (bottom row) - see table$\,$\ref{tab:sims}. The plots on the right show lineouts of the laser propagation axis at $y=0$. The laser has propagated $\sim\SI{0.7}{\milli\meter}$ within the plasma.} 
\end{figure}

The longitudinal electric field exhibits asymmetries when introducing clusters to the plasma (see Figs.$\,$\ref{fig:Exc1} and \ref{fig:Exc5}). The left hand side of the figure shows simulation data from three different simulations (D,E,F) at a point at which the laser has propagated $\sim\SI{0.3}{\milli\meter}$ within the plasma. It can be seen that the well known structure of a nonlinear wakefield as seen in the top panel is not maintained in the cluster cases and that an increase in cluster density (which also means that the inter-cluster separation is greater in order to keep the average density constant) reinforces this effect as shown in the bottom panel. High frequency oscillations can be observed in the cluster simulations (second and third row) which are due to the fields generated by the separation of electrons from their respective ion clusters. As in the density plots, the cluster structure is shown to decay behind the first wake when clusters are introduced. 

After the laser pulse has propagated over $\SI{1}{\milli\meter}$ in a cluster plasma, a significant amount of charge can be observed to be trapped within the first bubble (see Fig.$\,$\ref{fig:stream}) of a cluster simulation. The volume taken up by ions (shown in red) increases strongly towards the back (left) of the simulation box due to Coulomb forces experienced within the ion clusters (``Coulomb explosion''). The gradient depends on the charge contained in the cluster: higher peak ion charge density increases the expansion which scales with increasing Coulomb force $F_c \propto n_e\times r_{cl}$, where $r_{cl}$ is the cluster radius. Due to the random nature of the cluster center positions, regions of high charge fluctuations are present within the simulation enabling the aforementioned streams of electrons through the bubble region.

When encountering the laser pulse, the electrons expand on a different scale to the ions. Even though the ions strongly attract the electrons of the micro-plasma within the cluster structure, the forces caused by the laser electromagnetic field are strong enough to cause a dissociation of the electrons from the ion cluster over a short amount of time. Over this short period, the (electron) cluster expansion undergoes two different stages as can be seen in the lower panels of Fig.$\,$\ref{fig:stream}. In the region in which the laser intensity is very low the clusters undergo hydrodynamic expansion driven by small oscillations of the electrons about the ion cluster centre (see bottom left panel in Fig.$\,$\ref{fig:stream}). When the laser field increases electrons are ripped out of the ion cluster. First, the ponderomotive force pushes the entirety of the electron micro-plasma to a position where the force is balanced by the attracting ion charge. Due to the time spent outside of the ion cluster and the strong repulsive Coulomb force between the electrons, the micro-plasma is forced to expand (on a time scale much shorter than the expansion of ions of higher mass). For the electrons it holds that the larger the cluster diameter, the longer the electrons stay associated with the respective ion cluster. On the other hand, as soon as the charges are separated, just like in the ion case, the Coulomb explosion is stronger due to the larger amount of charge within the cluster.

\subsection{Electron Trapping and Acceleration}\label{sec:acceleration}
Higher self-injected electron charge, higher maximum energy, and a larger energy spread are three features that have been observed in experiments using clustered media targets as mentioned in the introduction to this article. Here, we show that the way electrons are trapped in the wakefield differs fundamentally from the uniform ``self injection" case. We will show that this new process leads to a large spread in electron energy and link the observations to cluster and laser parameters.

\begin{figure}
\subfigure{\includegraphics[width=1.0\linewidth]{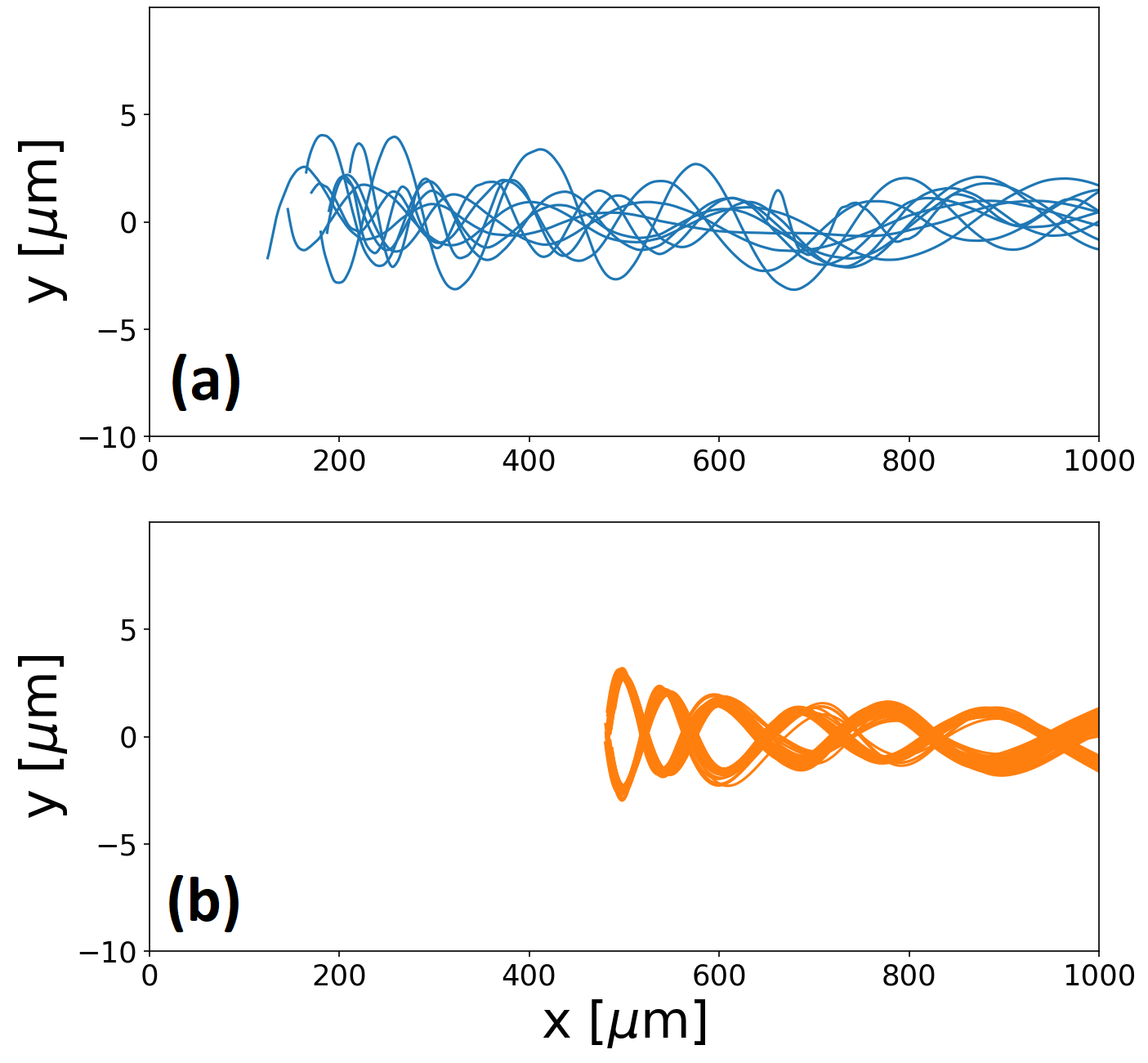}}
\subfigure{\includegraphics[width=1.0\linewidth]{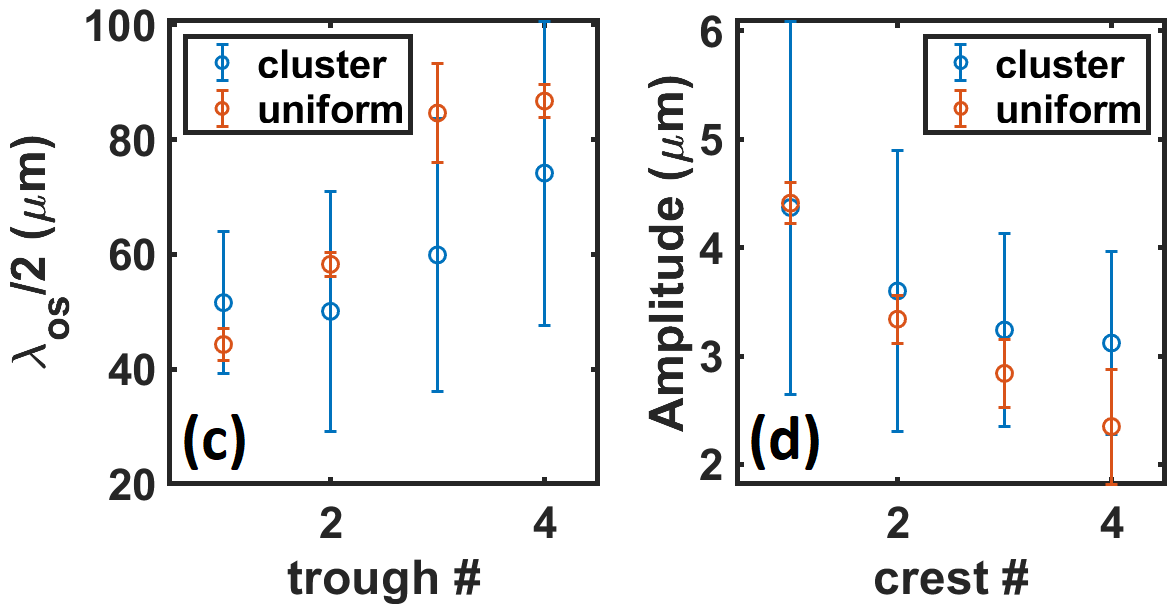}}
\caption{This figure presents particle trajectories of trapped electrons from the point of injection into the wake for the cluster (a) and uniform (b) case. Out of 100 trajectories, 10 tracks were randomly chosen and analysed (c,d). Half-wavelength distances and oscillation amplitude are plotted against the troughs (and crests) of the electron trajectory. The error bars correspond to the standard deviation (1$\sigma$) obtained from the 10 randomly chosen tracks.}
\label{fig:trajectory}
\end{figure}

\begin{figure*}
\subfigure{\label{fig:track2}\includegraphics[width=1.0\linewidth]{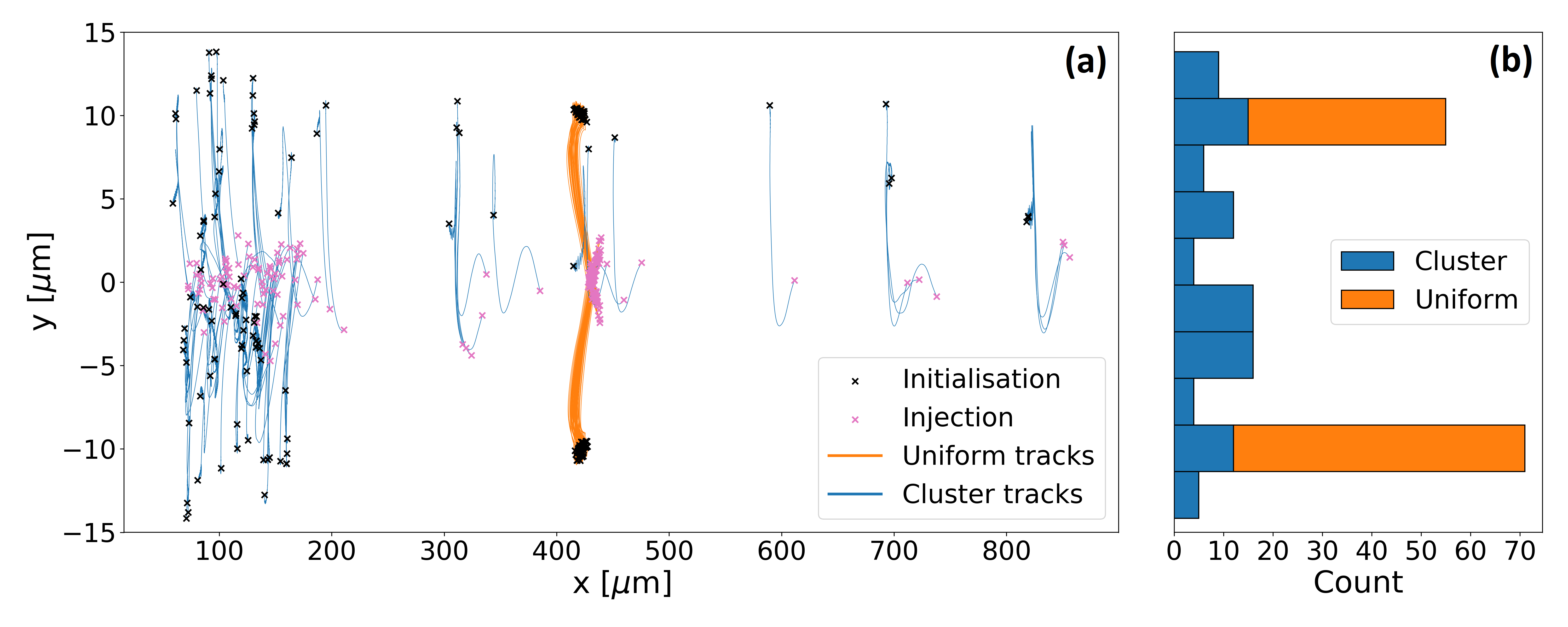}}
\subfigure{\label{fig:track3}\includegraphics[width=0.49\linewidth]{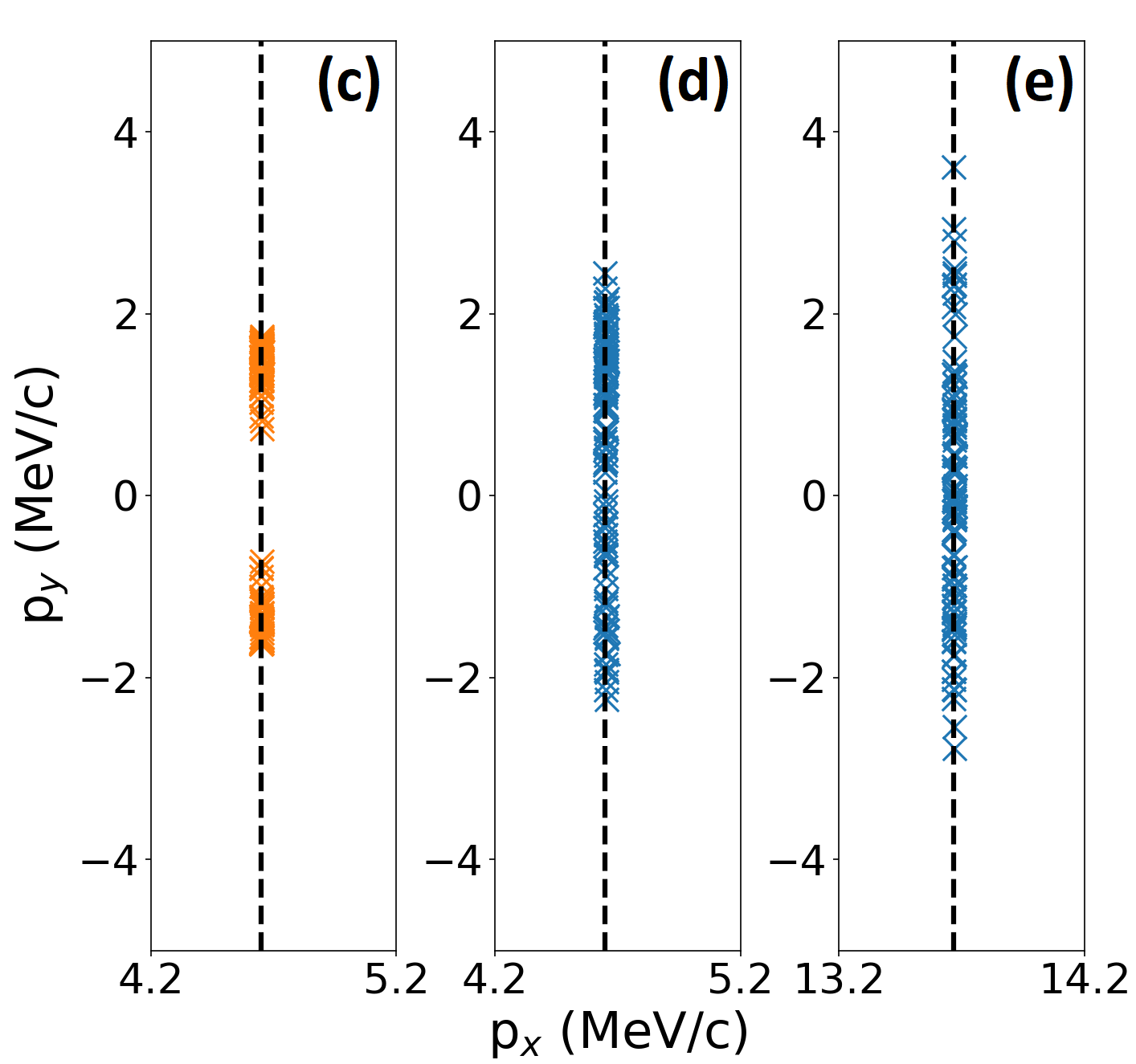}}
\subfigure{\label{fig:track4}\includegraphics[width=0.49\linewidth]{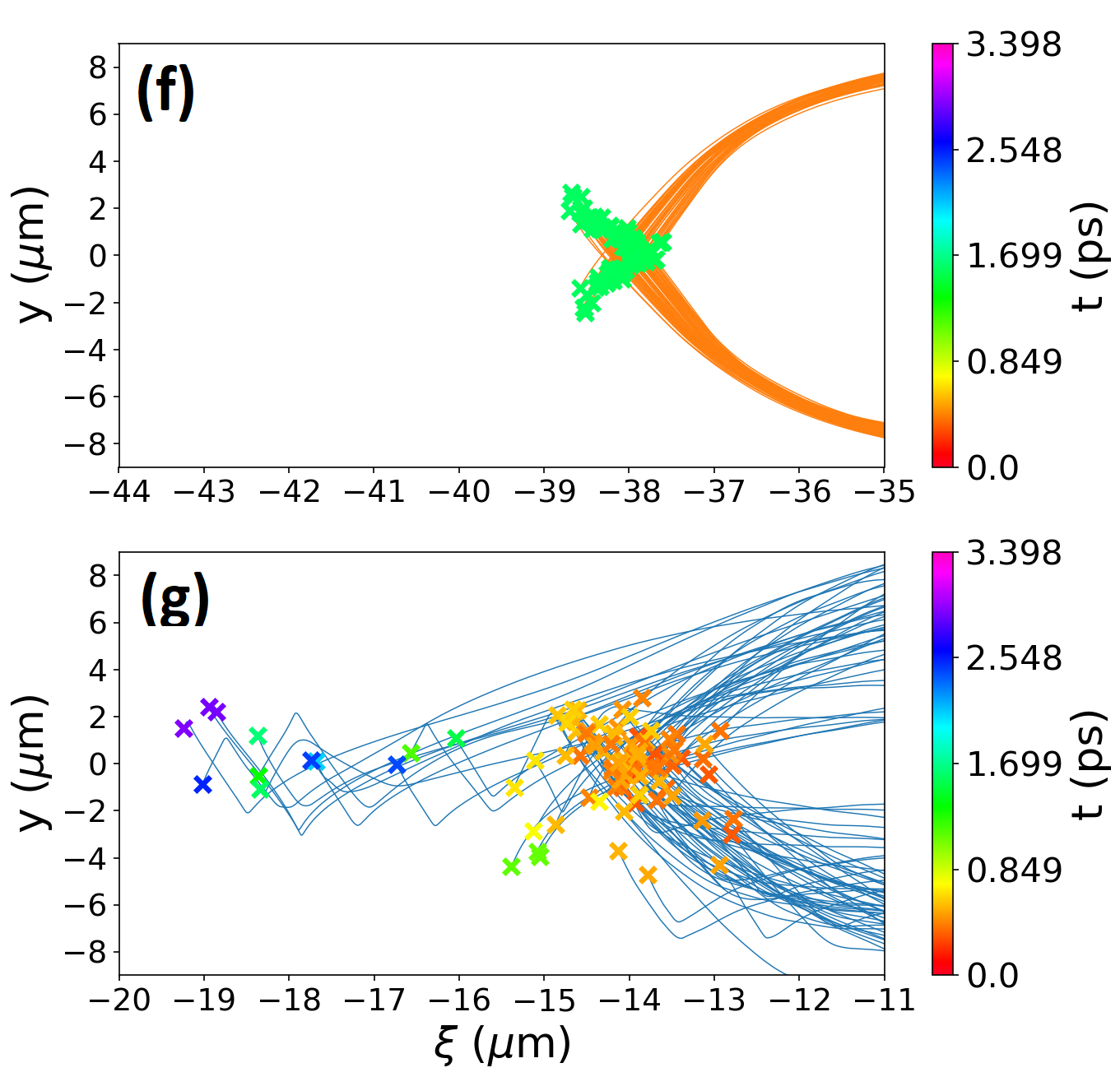}}
\caption{Analysis of the electron injection behavior. Positions of particle injection and corresponding initialisation (origin) positions for uniform and cluster plasmas (simulations D and G) are shown in panel (a). The histogram in panel (b) presents the abundance of particles born at a certain y-position (initialization at right border of simulation box).
The perpendicular momentum at time of injection within the wakefield (threshold calculated using eq.$\,$\ref{eq:Lu_inj}) is shown for the uniform (c) and cluster (d) case. Panel (e) gives the momentum for a threshold calculated using the higher laser group velocity found in section$\,$\ref{sec:laser}. Plots (f) and (g) show a comparison of injection position in a co-moving frame travelling at the speed of light for the uniform and cluster case. The color of the 'x'-markers corresponds to the time at injection. }
\label{fig:injection_tracks}
\end{figure*}

Fig.$\,$\ref{fig:trajectory} shows trajectories of particles trapped in a cluster (panel a) and a uniform plasma (panel b). The particles whose trajectories are displayed were randomly selected from a set of macroparticles satisfying two conditions: Firstly, a minimum longitudinal momentum threshold had to be reached to separate accelerated electrons from those that were not accelerated. Secondly, particles were spatially selected to restrict the data to the first bubble into which particles were injected. 

Chen \etal $\,$\cite{Chen2013} observed a strong resonant coupling between tracked electrons and laser field leading to oscillations at the betatron frequency (frequency of electrons oscillating within the bubble to create synchrotron radiation \cite{Esarey2002}) - in strong contrast to electrons accelerated in a uniform plasma where the transverse elongation of the trajectories was shown to be much smaller. The transfer from transverse to longitudinal momentum was then linked to energy gain through direct laser acceleration observing strong resonant coupling to the laser field. From our simulations, it can be seen that the mean amplitude and frequency of the oscillation is similar in both cases contrasting the findings in \cite{Chen2013}. The wavelengths of the oscillation are also similar in the cluster and uniform simulations with the electrons in the cluster case having a phase shift due to the different points of injection (see below).

The distance over which particles are injected is very narrow (localized injection) in the uniform case which means that the trapping condition (electrons within the wake reach a longitudinal momentum threshold that corresponds to a velocity higher than the wake's phase velocity $v_{\Phi}$) is met for a very short amount of time as can be seen in Fig.$\,$\ref{fig:track2}. In contrast, electron injection in the cluster regime occurs over a long distance which means that there has to be a source aiding injection. The cluster injection starts at an early stage - the first particle is injected roughly at the point where the density ramp has reached the bulk average density value of $\langle n_e\rangle=0.004\,n_{cr}$. It has to be noted that there is a selection bias towards early time steps due to the nature of the momentum cut-off meaning that more particles will reach the threshold if they were injected earlier adding a weight to the random selection.
The position on the y-axis at which trapped electrons enter the simulation box (initialized from the right-hand-side boundary) shows another important difference between uniform and cluster wakefields. It was shown in \cite{Kostyukov2004} that particles are self-injected from an annulus with a radius on the order of the laser spot size. This is also observed in the uniform plasma simulations (see histogram in Fig.$\,$\ref{fig:injection_tracks}). In the cluster case, however, particles can be injected from a great variety of positions even very close to the central axis. Electrons bound to a cluster can stay at a central position for a longer time similar to what happens in the ionization injection mechanism where particles are initialized to be available for injection at a point closer to the center of the laser pulse. The wake potential difference between the point of ionization (which, in our case, corresponds to the point where the electrons stop being bound to a particular ion cluster and cease to be dominated by the cluster's attracting fields which, again, depend on the cluster density and size) and the point of trapping was found (\cite{Pak2010}) to be $\Delta\Bar{\Psi}= \sqrt{1+(P_{\perp}/mc)^2}/\gamma_\phi-1$, where $P_{\perp}$ is the perpendicular electron momentum, and $\gamma_\phi = (1-(v_\phi/c)^2)^{-1/2}$. When electrons stay associated with the cluster for longer they are more likely to feel a greater potential difference. What adds to the effect of enhanced and continuous injection is the presence of ion clusters. As was shown in \cite{Cho2018} the potential of an ion nanoparticle injected into the wakefield eases the trapping condition. It can be argued that the same mechanism aids injection in our case (albeit to a smaller extent due to the smaller cluster size). Rather than having one large cluster potential the bubble region consists of hundreds of ion clusters that stochastically add small increments of momentum to (non-evacuated) electrons when they are in the vicinity of each potential.

When studying how the perpendicular electron momentum behaves during the injection process a similar picture to what is found in the initialisation position is observed. Using the equation for the wakefield phase velocity \begin{equation}\label{eq:Lu_inj}
    v_{\Phi}=v_g-v_{etch}\simeq c\sqrt{1-\left(\frac{\omega_p}{\omega_0}\right)^2}-c\left(\frac{\omega_p}{\omega_0}\right)^2
\end{equation} found in \cite{Lu2007} to estimate the injection threshold, we find that there is an absence of particles with perpendicular momentum close to zero in the uniform case showing that particles are injected from a path on which they experience symmetric forces about the axis of laser propagation and that all electrons are forced onto a similar betatron oscillation pattern. In contrast, in the cluster case the region around $p_y=0$ is well populated, even if using a modified wakefield phase velocity to compute the injection threshold (it will be shown in section \ref{sec:laser} that there is an increase in wakefield phase velocity in the cluster case). This indicates that there are particles being trapped that have experienced equal positive and negative perpendicular impulse. In the setting of wakefields this is only possible if the electrons either traverse through the wakefield via a central path close to the laser propagation axis whilst gaining longitudinal momentum (i.e. through a slingshot mechanism in the vicinity of a cluster as discussed in \cite{Cho2018}) or via a path that exhibits non-symmetrical, fluctuating electric fields which have been observed in this study. The great variety of available perpendicular momenta at injection also gives an explanation for the high standard deviation in oscillation amplitude in Fig.$\,$\ref{fig:trajectory}.

Another observation from our simulations can be made when looking at a ``co-moving" frame propagating at $c$. The bottom right panels in Fig.$\,$\ref{fig:injection_tracks} show the position relative to the moving-window simulation box at which particles are trapped. \cite{Lu2007} show that the acceleration mechanism is very stable. This cannot be said for the cluster case. The large distance between the leftmost and rightmost injected particle in the cluster case shows that the particles are trapped in a wakefield phase front that fluctuates over time making it easier to satisfy the trapping condition mentioned above (similar to the injection technique used in a density down-ramp). Looking at the time of injection, a certain trend can be seen: there is a shift towards the left hand side of the box with increasing time showing that the wakefield phase velocity is indeed evolving at a speed $v_{\Phi}<c$. However, the fluctuations within a certain time period are observed to be significant (i.e. oscillation of green and blue markers following the co-moving spatial variable $\xi$). 

\begin{figure*}
    \centering
    \includegraphics[width=\linewidth]{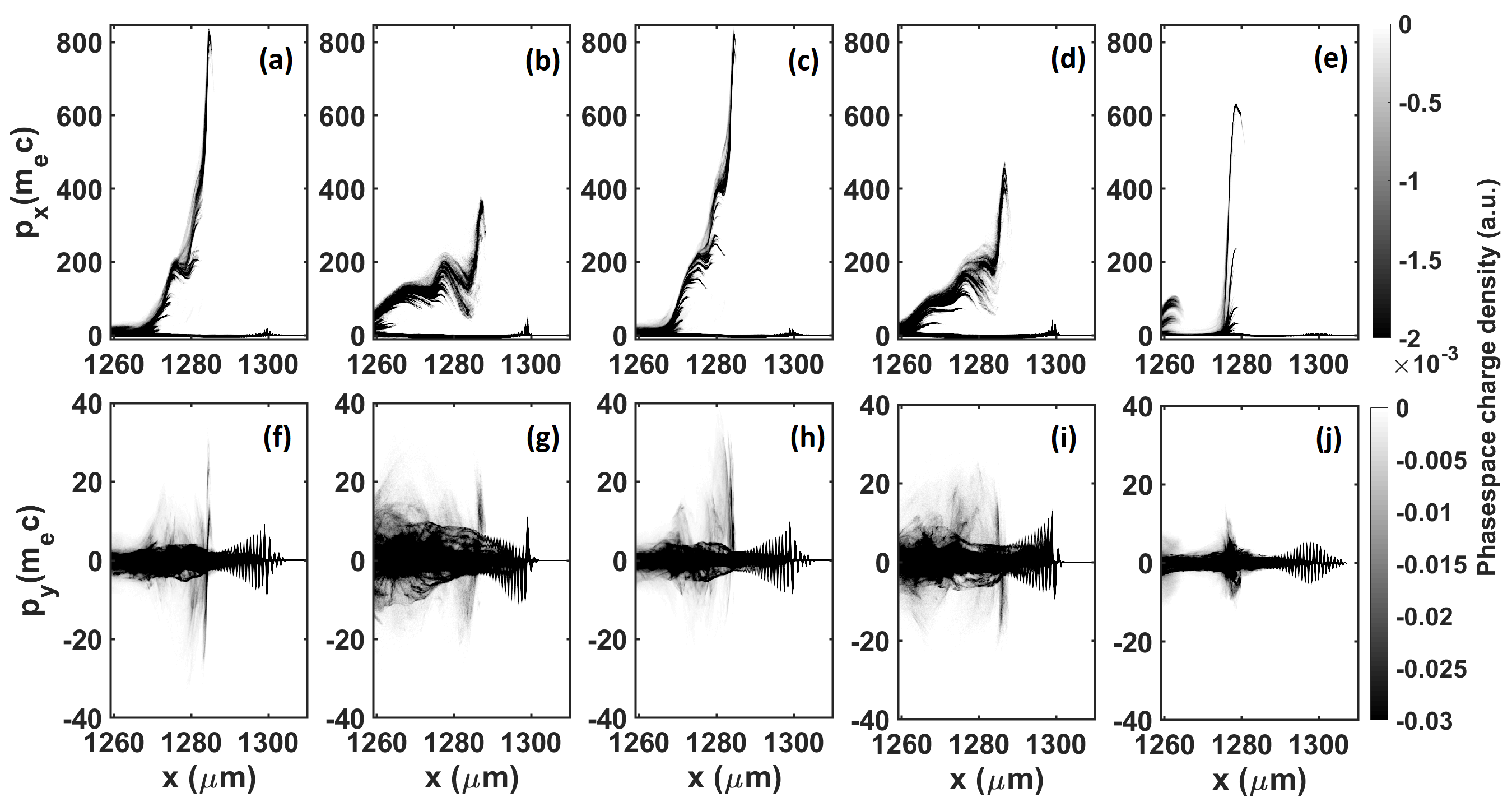}
    \caption{This figure shows the longitudinal and perpendicular momentum plotted against x-direction of cluster simulations with different cluster parameters. The data was taken after $\SI{4.33}{\pico\second}$. Panels (a) and (f) correspond to simulation E, (b) and (g) to F, (c) and (h) to G, (d) and (i) to H, (e) and (j) to J.}
    \label{fig:phasespace}
\end{figure*}

When looking at the phase spaces of the various simulations studied in this article an argument can be made for why different experiments observed different values for charge within the accelerated particle beam and maximum particle energy while consistently observing a broadened spectrum. Fig.$\,$\ref{fig:phasespace} shows the longitudinal momentum of the electrons within the simulation box for simulations E, F, G, H, and J (scan of cluster parameters according to Tab.$\,$\ref{tab:sims}). The data shows that in a cluster plasma electrons are accelerated in a very distinct manner in the sense that a spatially spread out bunch is observed that is constantly ``fed'' from the back of the wakefield bubble. The cluster diameter and peak density are observed to have an impact on the maximum momentum and charge distribution within the beam. The simulations using high density (b) and large diameter clusters (d) have almost identical cluster number density and, hence, cluster separation. The large cluster separation leads to a strongly fluctuating wakefield wave front that may overtake the formerly trapped particles. It can be assumed that this leads to kinks within the $p_x$ phasespace plots. The maximum momentum depends on the field strength. We have seen earlier that a plasma consisting of more, but smaller or lower-density clusters exhibits a smoother field structure and higher peak accelerating fields explaining the higher momentum for simulations E, G, and J.  
When looking at the perpendicular momentum $p_y$ we can infer from the electron response that the laser is etched to a greater extent in simulations F and H. The smallest etching effect is seen in the simulation using $\langle n_e\rangle/n_{cr}=0.002$, as is expected.
 
\begin{figure}
    \centering
    \includegraphics[width=\linewidth]{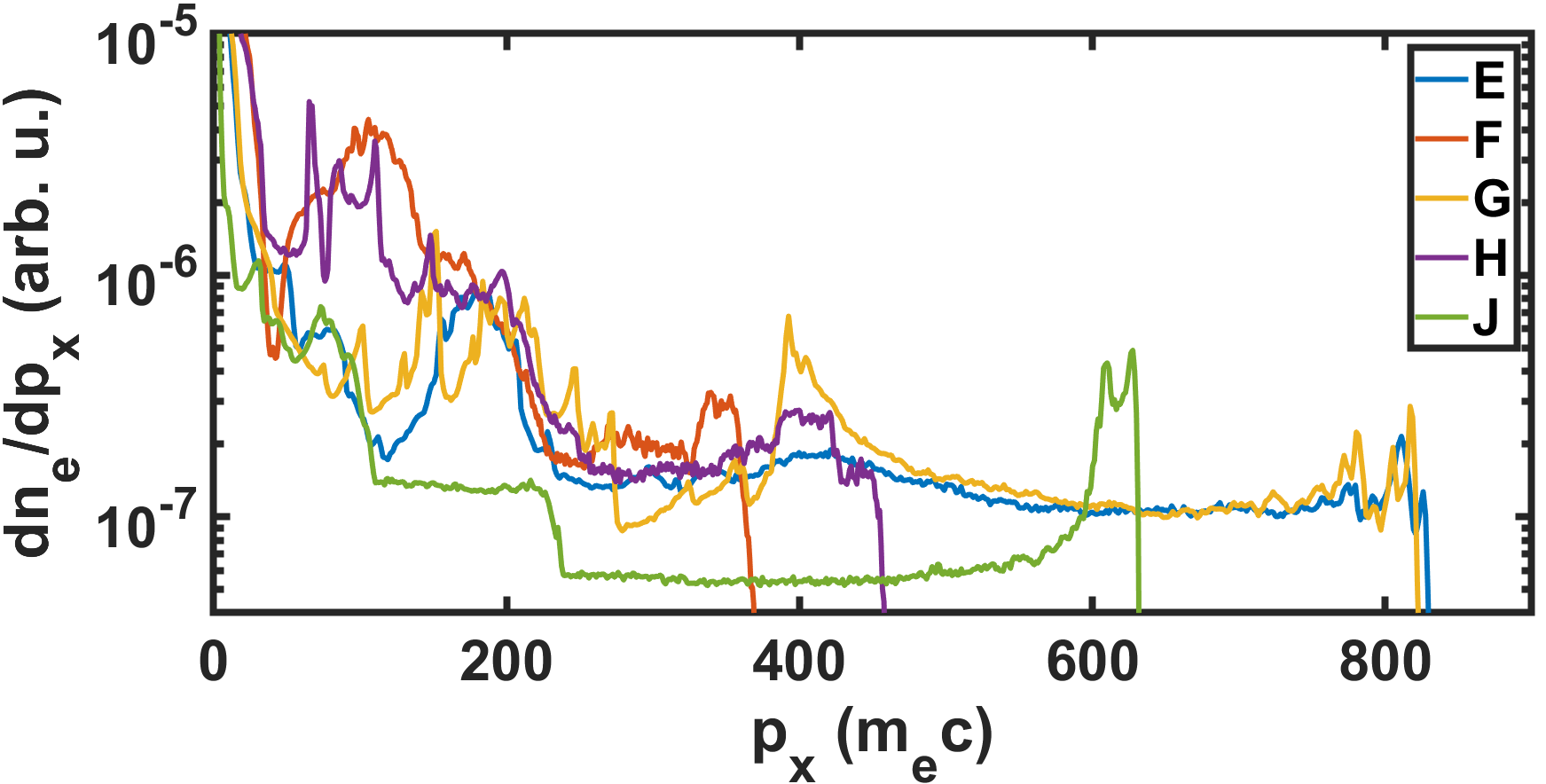}
    \caption{This figure shows the integrated charge within the simulation box against the longitudinal momentum for simulations using different laser and cluster parameters. All data was taken after $\SI{4.33}{\pico\second}$}
    \label{fig:spectrum}
\end{figure}

Plotting the spectrum (Fig.$\,$\ref{fig:spectrum}) shows that wakefields driven in plasmas of larger clusters show a decrease in maximum electron energy compared to those driven in plasmas of small clusters. Simulations E and G show almost identical acceleration spectra above a certain threshold. Reducing the average density by a factor two (J) creates a distinguished peak making the spectrum of the accelerated bunch closer to mono-energetic. The smallest acceleration is experienced by trapped electrons in a plasma of dense clusters (F). There is an increase of maximum momentum with increasing cluster number density (with the exception of saturation for very high cluster numbers where E and G have the same maximum momentum). A similar (inverse) correlation can be found between cluster peak electron density and maximum momentum with the same exception for E and G. Of course, these correlations also hold for when replacing the maximum momentum with the magnitude of the accelerating field.

The amount of charge accelerated in the electron beam after $\SI{4.33}{\pico\second}$ above a threshold of $80\%$ of the respective maximum momentum and above $500\,\mathrm{m_ec}$ is shown in Tab.$\,$\ref{tab:charge}. 

\begin{table}[h!]
\centering
 \begin{tabular}{||c|c| c| c| c|c|c|c|c|c||} 
 \hline
 Simulation & D & E & F & G & H & I & J & K & L \\ [1ex] 
 \hline
  Q ($>80\%$)  & 8.0$^*$  & 7.2 & 1.6 & 5.4 & 9.1 & 15.2 & 7.6 & 6.0* & 9.6 \\ [1ex] 
  \hline
  Q ($>500\,m_ec$) & 0.0 & 22.1 & 0.0 & 18.1 & 0.0 & 0.0 & 7.9 & 0.0 & 12.8 \\ [1ex] 
 \hline
 \end{tabular}
\caption{List of charge accelerated within the highest $20\%$ of momentum and above $500\,\mathrm{m_ec}$ for different simulations. All values are in $\si{\pico\coulomb}$. The asterisk indicates that the value was calculated from an earlier time step in simulations D and K compared to all other simulations due to dephasing and charge reduction of electrons at later time steps.}
\label{tab:charge}
\end{table}

The two dimensional nature of the simulation data prevents a direct read-out of the beam charge from the available diagnostics. To get an estimate, we isolated background electrons using a $p_x$ threshold below which data was ignored. 
Taking the integrals of the charge density and its square along the second (transverse) dimension for a given x-cell yields
\begin{equation}
    \int{n_e dy} \sim 2 n_0 w_0\ ,
    \end{equation}
    \begin{equation}
    \int{n_e^2 dy} \sim 2 n_0^2 w_0\ .
\end{equation}
Here, and in what follows ``$\sim$" becomes ``$=$" for a top hat beam profile with half-width $w_0$. We can rewrite $w_0$ as
\begin{equation}
    w_0 \sim \frac{\left(\int{n_e dy}\right)^2}{2\int{n_e^2 dy}} = \frac{4n_0^2w_0^2}{4n_0^2w_0}\ .
\end{equation}
The total amount of particles within a two-dimensional structure in y-z-space that can be approximated using a cylindrically symmetric distribution function is
\begin{equation}\label{eq:N}
    \begin{split}
    N dx \sim \pi w_0^2n_0 dx &= 2n_0w_0\frac{\pi}{2}w_0 dx \\
    &= \int{n_e dy} \frac{\pi}{4}\frac{\left(\int{n_e dy}\right)^2}{\int{n_e^2 dy}} dx \\
    &= \frac{\pi}{4}\frac{\left(\int{n_e dy}\right)^3}{\int{n_e^2 dy}} dx\ . \\
    \end{split}
\end{equation}
This calculation can be repeated for every position in x-space. Integrating over $dx$ gives the total amount of particles within the beam independent of the actual shape of the structure. Errors arise from the approximation of every $dx$ value being cylindrically symmetric. However, no assumption on the symmetry of the total beam was made. It is important to note that the result is independent of the parametrization of the density profile, beam width and shifts from the centre.
The values for a Gaussian distribution can be readily obtained replacing the factor $\pi/4$ in the last line of Eq.$\,$\ref{eq:N} with $1/\sqrt{2}$ which gives a difference of $10\%$. This means that the value is correct within $10\%$ for any super-Gaussian distribution at each x-slice within the simulation box. 
Another error may arise if there is more than one bunch co-propagating at separation distance $s$. If there is no overlap ($s\gg w_0$) then, for two Gaussian beams the charge has to be multiplied by a factor $f=1/2$. If there is perfect overlap ($s=0$), the charge is accounted for accurately. In between, the factor is found to be \begin{equation}
    f = \frac{1+e^{-\frac{1}{2}(s/w_0)^2}}{2}\ .
\end{equation} 


\subsection{Analysis of Laser Pulse Properties}\label{sec:laser}
Tajima \etal show \cite{tajima2} that the dispersion relation of electromagnetic waves in a cluster plasma is different to that in a uniform plasma because of a restoring force due to the polarizability of the cluster: 
\begin{equation} \label{eq:tajima}
n^2(k,\omega)\equiv \frac{c^2k^2}{\omega^2}=1-\frac{p\omega^2_p}{\omega^2-f\omega^2_p+i\gamma\omega}-\frac{q\omega^2_p}{\omega^2}\ .
\end{equation}
Here, p is the "packing" ratio of cluster volume to entire volume, f is a factor of order unity related to the geometry of the cluster, $\gamma$ is the damping rate, and q is the ratio between the background plasma density and the cluster density. The density corresponding to $\omega_p$ in this equation is the density within the cluster.
This leads to the above mentioned cluster mode that allows a laser pulse to propagate in a plasma of highly overcritical clusters.

The propagation of a laser pulse in a plasma of overcritical clusters corresponding to Eq.$\,$\ref{eq:tajima} was tested using two-dimensional particle-in-cell simulations (see Tab.$\,$\ref{tab:sims}). However, it has to be mentioned that this theory of cluster polarization inevitably breaks down as soon as the clusters are disassembled and electrons are not associated with a specific ion cluster anymore. At the intensities used in our simulations cluster disassembly starts early on in the laser-cluster interaction but depends on the cluster parameters. Hence, one would expect that the impact of the polarization effect is stronger in simulations using clusters of higher density and size at the front of the laser because the electrons stay associated with the respective ion cluster for a longer time (at the back of the laser, the high charge will lead to fast expansion through the Coulomb explosion mechanism). However, we find that the deterioration of the laser shape makes it difficult to give an accurate account of what the group velocity is for high density clusters.
A magnitude that can be measured, however, is the maximum peak velocity which may shift forwards in the moving-window box. Taking a lineout through the centre of the pulse along the laser propagation direction and measuring the position of the wave packet central peak at initialization ($t=0$) gives a reference value. Measuring the peak position caused by the bunching of red-shifted photons at the front of the pulse at the final time step ($t=\SI{4.33}{\pico\second}$) and dividing by the simulation time, we find a maximum-peak velocity of $v_{mp} = 1.0004\,c$ (simulation G) and $v_{mp} = 0.9993\,c$ (simulation D). Comparing the difference of these two respective measurements indicates that either the etching velocity in the cluster case is lower or the group velocity is higher, both impacting the wakefield phase velocity. Analysing other cluster simulations shows that, in the high intensity regime, the group velocity of the laser pulse does not seem to depend strongly on cluster parameters for the parameters studied in this paper.

This result is important in showing that using cluster targets a slight increase in dephasing length $L_{dp}$ of the plasma accelerator ($L_{dp} \simeq (1-(v_{g}-v_{\mathrm{etch}})/c)^{-1} R$, where $v_g$ is the laser group velocity, $v_\mathrm{etch}$ is the laser etching velocity, and $R$ is the distance between the back and the center of the wakefield) can be achieved. As shown above, $R$ fluctuates in the cluster case which means that some trapped electrons experience an increase in dephasing length leading to a higher peak energy. Additionally, on average, the increase in group velocity will add to the effect of increasing the dephasing length.  
A change in wakefield phase velocity $v_{\Phi} \simeq v_{g}-v_{\mathrm{etch}}$ from $v_{\phi}^{\mathrm{unif}}$ to $v_{\phi}^{\mathrm{cl}}$ leads to an extension of the dephasing length by a factor \begin{equation}
    \frac{L_{dp}^\mathrm{cl}}{L_{dp}^\mathrm{unif}}=\frac{1-v_{\phi}^{\mathrm{unif}}}{1-v_{\phi}^{\mathrm{cl}}}
\end{equation} increasing the final maximum energy output.   


Laser etching was studied by Decker \etal \cite{Decker1996}. It was found that the velocity at which the the front of the laser pulse is etched backwards is $v_\mathrm{etch}\propto \omega_p^2/\omega_0^2$ which means that the laser will be depleted after a distance of $L_{etch}\simeq c\tau_\mathrm{FWHM}\omega_p^2/\omega_0^2$. The frequency shift of a laser pulse in a plasma density gradient was described in \cite{Wilks1989} and experimentally observed in \cite{Murphy2006}. When a wakefield is driven by a short laser pulse a red-shift can be observed. This effect is also called ``photon deceleration" and is associated with the energy loss from driving the wakefield whilst preserving the photon number within the pulse \cite{Chen1999,Spitkovsky2001}. Here, we will show how this process scales with cluster parameters. 

\begin{figure}
    \includegraphics[width=\linewidth]{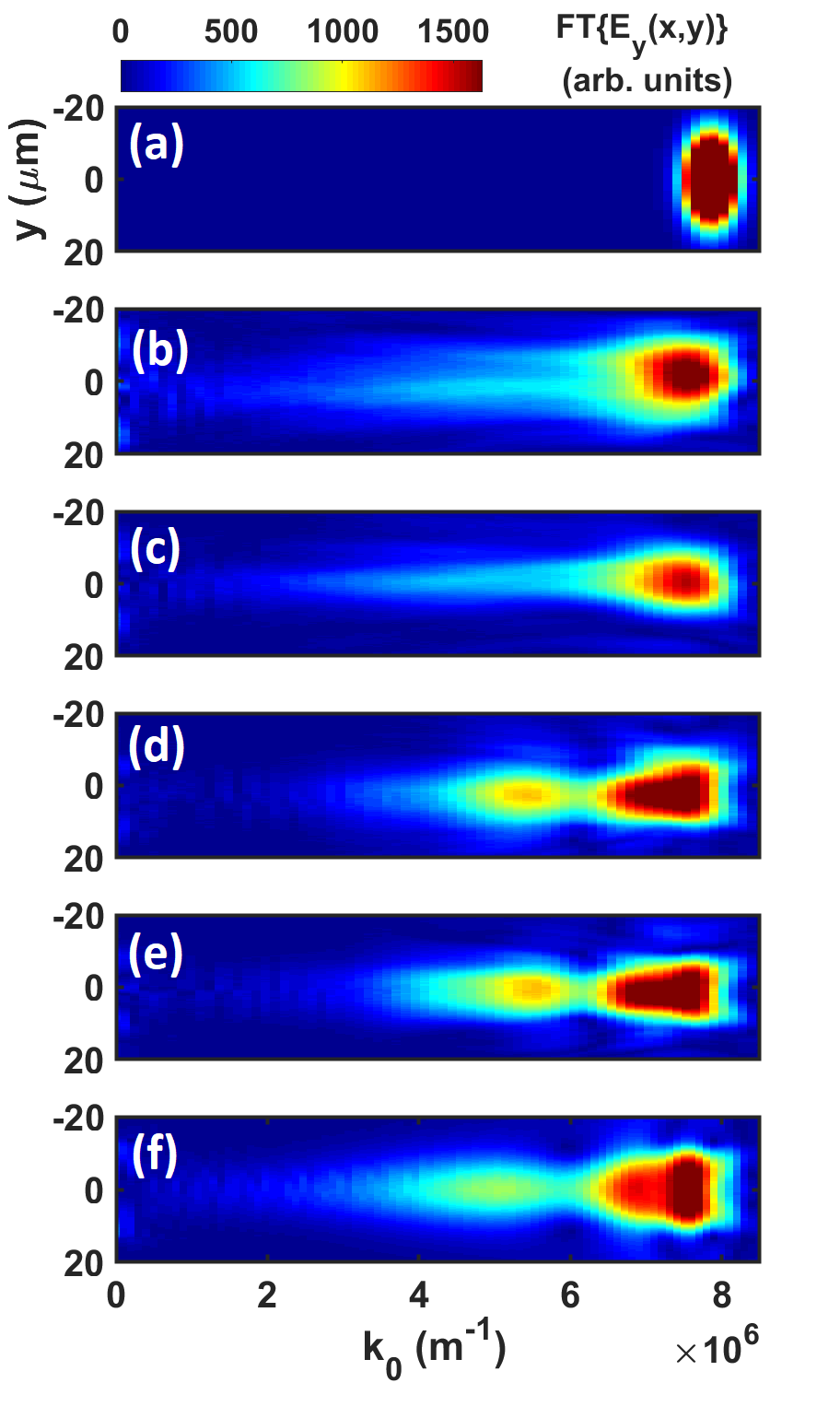}
    \caption{Fourier transform of $E_y(x,y)$ field as a function of wave vector $k_0$ and $y$ at initialization (a) and after propagating $\SI{4.33}{\pico\second}$ (b-f). Panels (b-e) show results using different cluster parameters and panel (f) shows data from a uniform simulation. The cluster peak density (normalized to $n_{cr}$) is 5 (b), 2.5 (c), 1.25 (d), and 0.75 (e). The relative normalized number density of clusters in the simulation box is $0.15$ (b), $0.17$ (c), $0.6$ (d), and $1$ (e) giving the same average electron charge density. The wave vector for a laser of wavelength $\lambda_0 = \SI{800}{\nano\meter}$ is $k_0 = 7.85\times 10^6\si{\per\meter}$ in vacuum and $k_0 = 7.84\times 10^6\si{\per\meter}$ in a plasma of $n_e=0.004\, n_{cr}$.}
    \label{fig:redshift}
\end{figure}

\begin{figure}
    \includegraphics[width=\linewidth]{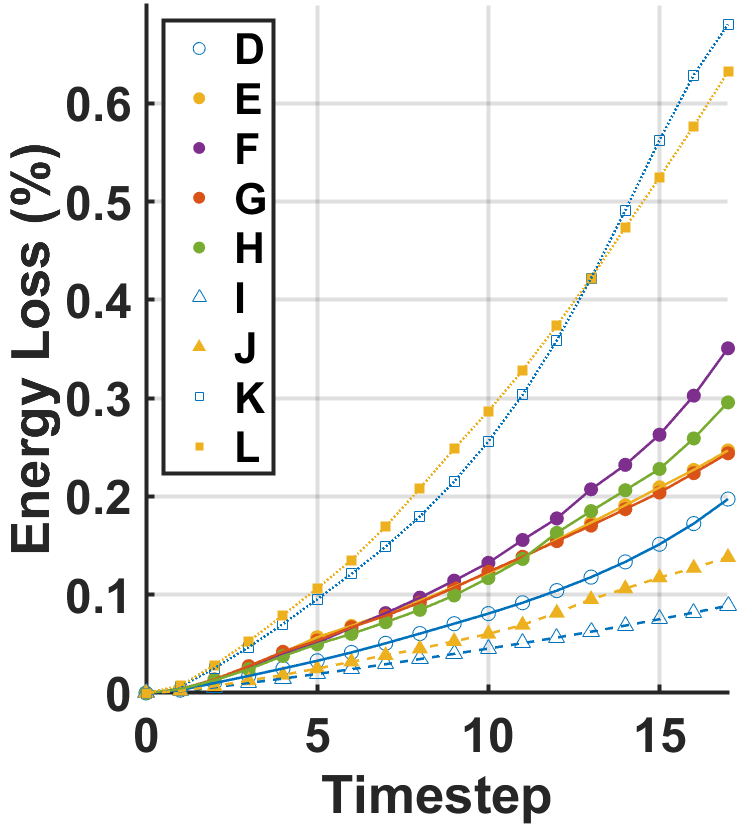}
    \caption{Energy loss normalized to the laser pulse energy at initialization plotted over time. Hollow marker objects indicate the use of a uniform plasma, filled marker objects indicate a cluster plasma. Triangles represent simulations using half the plasma density ($0.002\,n_{cr}$). Squares represent the use of laser pulses of only $18.9\%$ of the energy (compared to all other simulations) within a halved pulse length of ($\SI{15}{\femto\second}$) and reduced spot-size. The parameters of each simulation can be found in table$\,$\ref{tab:sims}. Here, one time step corresponds to $\SI{0.25}{\pico\second}$ in simulation D and to $\SI{0.26}{\pico\second}$ in all other simulations.}
    \label{fig:energy_loss}
\end{figure}

We have seen in Fig.$\,$\ref{fig:phasespace} that the simulations using denser (panel g) and larger (panel i) clusters show a stronger impact on the depletion of the front of the laser pulse but not on the etching velocity itself. Taking the Fourier transform of the perpendicular electric field $E_y(x,y)$ of the laser pulse shows (see Fig.$\,$\ref{fig:redshift}) that the use of dense (panel b) and large (panel c) clusters leads to a lesser degree to a red-shifted spectrum than the less dense cluster simulations (panels d, e) and the uniform simulation (panel f). This indicates that a wakefield is efficiently driven in simulations D, E, and G, whereas energy must be lost through a different mechanism in simulations F and H. We argue that, as was discussed in \cite{fennel}, the stronger impact of the effect of resonant laser absorption for high density clusters along with photon scattering from overcritical targets causes the strong etching described above and depletes the laser without driving a wakefield efficiently. To support this claim, we have measured the energy loss over time for all simulation parameters. The results are shown in Fig.$\,$\ref{fig:energy_loss}. It can be seen that the laser pulse loses more energy in simulations F and H (compared to simulations E and G), even though no strong red-shifted peak is observed. We find strong relative laser depletion when using a laser pulse of lower initial energy for both uniform (K) and cluster (L) cases. When reducing the average plasma density by a factor two but keeping the laser parameters unchanged (I, J), we find that less energy is lost in both cases, as expected.

\section{CONCLUSION}
To conclude, we presented the first high-resolution numerical study of the interaction of a high-power laser pulse with a cluster plasma in which it was shown that it is important to simulate cluster plasmas with random cluster positions rather than periodically aligned positions. Our data shows that there is a new injection mechanism in cluster plasmas, acting in a similar way to ionisation injection. It occurs due to the laser ripping electrons off the clusters, which then become trapped in the wake. Resulting from this process, electrons can be continuously trapped due to fluctuations in the bubble radius/shape caused by the random distribution of cluster positions and it was found that trapped electrons may originate from a position close to the laser propagation axis. This continuous injection mechanism is linked to a large spread in longitudinal and perpendicular momentum which is found across a range of cluster parameters and explains the large emittance found in experimental investigations elsewhere. We found that the electron bunch charge is a function of the cluster parameters which makes it able to tune x-ray betatron emission by changing cluster parameters. Other parameters influencing betatron x-ray flux such as higher perpendicular momentum are also found to be increased. In contrast to other numerical studies, we found that the betatron oscillation frequency and amplitude are of the same order for cluster and uniform plasmas. Finally, we have shown that the efficiency with which the laser pulse drives the wakefield depends on cluster parameters and that an increase in dephasing length could be achieved using cluster targets. 

\begin{acknowledgments}
One of the authors (M.W.M.) would like to thank Anton Helm for useful discussions. This work has been funded by EPSRC grant numbers EP/L000237/1 and EP/R029148/1, and by STFC grant numbers ST/M007375/1 and ST/P002048/1. This research was also supported by the European Research Council (InPairs ERC-2015-AdG grant no. 695088) and the Leverhulme Trust. The authors would like to thank the UCLA/IST Osiris consortium. We gratefully acknowledge all of the staff of the Central Laser Facility and the Scientific Computing Department at STFC Rutherford Appleton Laboratory. This work used the ARCHER UK National Supercomputing Service (http://www.archer.ac.uk) and STFC’s SCARF cluster.
 
\end{acknowledgments}

\bibliography{refs}

\end{document}